\documentstyle[12pt]{article}
\begin{document}
\vskip .2 cm
\begin{center}
\LARGE {\bf A new "hidden colour hypothesis" in hadron physics} 
\vskip 3 cm 
{\bf Afsar Abbas} \\
Institute of Physics, Bhubaneswar-751005, India \\
(e-mail : afsar@iopb.res.in)
\vskip 40 mm
{\bf Abstract}
\end{center}


A new "Hidden Colour Hypothesis" within the framework of QCD, 
as an extension of and in keeping with the spirit of the 
"Colour Singlet Hypothesis" is hereby proposed 
As such it should play a role in a consistent 
description of exotic hadrons, like the diquonia, pentaquarks, dibaryons 
etc. How these exotic hadrons are affected by this new hypothesis is 
discussed here. This new hypothesis suggests that the experimentalists 
may not be looking for single exotics but for composites of two or 
more of the same.

\newpage

{\bf INTRODUCTION}

\vskip .5 cm

As per the colour confinement hypothesis of QCD only the colour singlet 
part of the quark configurations would manifest themselves as physically 
observed particles, All the other representations like the colour octet,
though they arise in qqq and q$\bar{q}$ systems, do not show up as 
physical particles. 
These forbidden configurations though show up in larger number 
multiquark systems to give colour singlet objects themselves. So such 
components though missing in simple qqq and q$\bar{q}$ systems actually 
play a prominent role in large 
number of quark colour singlet objects. So to say they are hidden inside 
larger colour singlet objects. As such these hidden colour parts ( to 
be clearly defined below ) are actually as significant as the colour 
singlet part. No one has been able to prove the colour singlet 
hypothesis from the fundamentals of the QCD. 
It may become a "Law" only after it can satisfactorily be shown to 
arise from QCD itself. So though no violation of it has 
been found ( remember that actually the colour singlet idea is first an 
empirical fact then later only a theoretical concept ) it still retains 
the prefix of "hypothesis".

Now to define the "hidden colour" concept. The baryons, made up of three 
quarks, besides the colour singlet part which shows up in nature 
explicitly as per the colour singlet hypothesis, have other forbidden 
parts like the colour octet. When one goes to the six quarks then the 
total composite system can be colour singlet in the following manner: 

\begin{equation}
| 6q > = \frac{1}{ \sqrt{5} } | SS > + \frac{2}{ \sqrt{5} } | CC >
\end{equation}

where S represents a 3-quark cluster which is singlet in colour space and
C represents the same as octet in colour space. Hence $ | CC > $ is
overall colour singlet. This part is called the hidden colour 
because as per confinement ideas of QCD these octets cannot be separated 
out asymptotically and so manifest themselves only within the 6-q 
colour-singlet system \cite{ref1}. 

\vskip .5 cm
{\bf SOME GROUP THEORY}
\vskip .5 cm

 Let $ S(f_{1})$ and $S(f_{2})$ be the permutation groups for the particles 1,
 2, ....$f_{1}$ and $f_{1}+1$, $f_{2}+2$, .... $f_{1} + f_{2}$, respectively,
 with $f= f_{1}+ f_{2}$. We use the following notation \cite{ref8} to
 designate the irreps of the nine groups $S^{x}(f_{1})..... S^{q}(f)$ and the
 irreps of SU(m), SU(n) and SU(mn) ( of the group chain $ SU(mn)\supset SU(m)
 \times SU(n)$)

\begin{equation}
 \left( \begin{array}{clc}
\sigma^\prime  & \mu^\prime  & \nu^\prime\\
\sigma^{\prime\prime} & \mu^{\prime\prime} & \nu^{\prime\prime}\\
\sigma      & \mu      & \nu
\end{array}      \right)  
 \left( \begin{array}{clc}
S^{x}(f_{1})   & S^{y}(f_{1})   & S^{q}(f_{1})\\
S^{x}(f_{2})   & S^{y}(f_{2})   & S^{q}(f_{2})\\
S^{x}(f)       & S^{y}(f)       & S^{q}(f)
\end{array}        \right) 
 \left( \begin{array}{clc}
SU(m) & SU(n) & SU(mn)\\
SU(m) & SU(n) & SU(mn)\\
SU(m) & SU(n) & SU(mn)
\end{array}       \right) 
\end{equation}
For example,  $\{\mu^{\prime\prime} \}$ and  $\{\sigma\}$ are the irreps 
of $S^{y}(f_{2})$ and $S^{x}(f)$ respectively. Let
\begin{equation}
  \left| \begin{array}{c}
\{\nu^\prime\}  \\
\beta^\prime \{\sigma^\prime\} W_{1}^\prime \{\mu^\prime\} W_{2}^\prime
\end{array}          \right\rangle ,
 \left| \begin{array}{c}
\{\nu^{\prime\prime}\}\\
\beta^{\prime\prime}\{\sigma^{\prime\prime}\} W_{1}^{\prime\prime} 
\{\mu^{\prime\prime}\}W_{2}^{\prime\prime}
\end{array}   \right\rangle ,
 \left| \begin{array}{c}
\{\nu\} \\
\beta\{\sigma\} W_{1}\{\mu\} W_{2}
\end{array}     \right\rangle 
\end{equation}
be the $SU(mn) \supset SU(m) \times SU(n)$ irreducible bases in the q-space
 of the particles $(1,2, \cdots f_{1})$, $(f+1, \cdots f)$ and
 $(1,2,\cdots f)$ respectively; $\{\sigma\}$, $\{\mu\}$  and $\{\nu\}$  are
 partition labels for the irreducible representations of SU(m), SU(n) and
 SU(mn) respectively; $W_{1}(W_{2})$ is the component index for the irrep
 $\{\sigma\}$ $(\{\mu\})$; and $\beta^\prime$, $\beta^{\prime\prime}$ and
 $\beta$ are inner multiplicity labels  \\ 
      $\beta^\prime$ = $ 1, 2,\cdots (\sigma^\prime \mu^\prime \nu^\prime)$,
 $\beta^{\prime\prime}= 1, 2, \cdots (\sigma^{\prime\prime} \mu^{\prime\prime}
 \nu^{\prime\prime}$), $\beta = 1, 2, \cdots (\sigma \mu \nu)$\\The integers
 ($\sigma^\prime\mu^\prime \nu^\prime$), ($\sigma^{\prime\prime}
 \mu^{\prime\prime} \nu^{\prime\prime}$) and ($\sigma \mu \nu$) are
 determined by the CG series of the permutation group. the $SU(mn)\supset
 SU(m)\times SU(n)$ CFP are defined as the coefficients in the fallowing
 expansion:

\begin{eqnarray}
\left| \begin{array}{c} \{\nu\} \tau\\ \beta\sigma W_{1}\mu W_{2} \end{array}  
\right\rangle 
&=& \sum_{\beta^\prime\beta^{\prime\prime}\sigma^\prime\sigma{\prime\prime}
\mu^\prime\mu^{\prime\prime}\theta\phi} 
C^{\{\nu\}\tau,\beta\{\sigma\}\theta\{\mu\}\phi}_{\nu^{\prime}\beta^{\prime}
\sigma^\prime
\mu^\prime,\nu^{\prime\prime}\beta^{\prime\prime}\sigma^{\prime\prime}
\mu^{\prime\prime}} \nonumber \\
&&\left[\left| \begin{array}{c} \{\nu^\prime\} \\ \beta^{'}\sigma^{'}\mu^{'}
 \end{array}
\right\rangle  
\left| \begin{array}{c} \{\nu^{''}\} \\ \beta^{''}\sigma^{''}\mu^{''} 
\end{array} \right\rangle\right]_{W_{1}W_{2}}^{\{\sigma\}\theta\{\mu\}\phi}
\end{eqnarray}
where $\tau$, $\theta$ and $\phi$ are the outer multiciplicity labels 
$ \tau = 1, 2,\cdots\{\nu^{'}\nu^{''}\nu\}$, $\theta = 1, 2,\cdots\{\sigma^{'}
\sigma^{''}\sigma\}$, $\phi = 1, 2,\cdots\{\mu^{'}\mu^{''}\mu\}$\\ \\
The integers $\{\nu^{'}\nu^{''}\nu\}$, $\{\sigma^{'}\sigma^{''}\sigma\}$ and
 $\{\mu^{'}\mu^{''}\mu\}$ are  decided by the Littlewood rule.\\
The bases are finally combined in to the irreducible basis $\{\sigma\}W_{1}$
 and$ \{\mu\} W_{2}$ in terms of the CG coefficients of $SU(m)$ and $SU(n)$
 respectively.\\ 
Next the irreducible bases classified according to the irreps of the group
 chain $S(f)\supset S(f_{1})\times S(f_{2})$ in the x,y and q(x,y) spaces be
 respectively denoted by
\begin{equation}
\label{abc}
 \left| \begin{array}{c}
\{\sigma\} \\
\theta\{\sigma^{'}\}m_{1}^{'}\{\sigma^{''}\}m_{1}^{''} 
\end{array}     \right\rangle ,
\nolinebreak
 \left| \begin{array}{c}
\{\mu\} \\
\phi\{\mu^{'}\}m_{2}^{'}\{\mu^{''}\}m_{2}^{''} 
\end{array}     \right\rangle , \left| \begin{array}{c}
\{\nu\} \\
\tau\{\nu^{'}\}m^{'}\{\nu{''}\}m^{''} 
\end{array}     \right\rangle 
\end{equation}

where  $m_{1}^{'}$, $m_{2}^{'}$ etc. are the Yamanouchi symbols. 
The $S(f)\supset S(f_{1})\times S(f_{2})$ ISF are defined to be  
the expansion coefficients in the fallowing:

\begin{eqnarray}
\left| \begin{array}{c}
\{\nu\}\beta \\
\tau\nu^{'}m^{'}\nu{''}m^{''} 
\end{array}     \right\rangle &=& \sum_{\beta^{'}\beta^{''}\sigma^{'}
\sigma{''}\mu^{'}\mu^{''}\theta\phi} 
C^{\{\nu\}\beta,\tau\{\nu^{'}\}\beta^{'}\{\nu^{''}\}\beta^{''}}
_{\{\sigma\}\theta\sigma^{'}\sigma^{''},\{\mu\}\phi\mu^{'}\mu^{''}}
\nonumber\\
&& \left[\left| \begin{array}{c}
\{\sigma\} \\
\theta\sigma^{'}\sigma^{''}
\end{array}     \right\rangle
 \left| \begin{array}{c}
\{\mu\} \\
\phi\mu^{'}\mu^{''}
\end{array}   \right\rangle \right]_{m^{'}m^{''}}^{\{\nu^{'}\}\beta^{'}
\{\nu^{''}\}\beta^{''}}
\end{eqnarray}
 
where the coupling in terms of the CG coefficients is indicated on the square
 bracket.
\\
it can be shown that the $SU(mn)\supset SU(m)\times SU(n)$ CFP
 are identical with the $S(f)\supset S(f_{1})\times SU(f_{2})$ ISF, i.e.
\begin{equation}
C^{\{\nu\}\tau,\beta\{\sigma\}\theta\{\mu\}\phi}_{\{\nu^{'}\}\beta^{'}
\sigma^{'}\mu^{'},\{\nu^{''}\}\beta^{''}\sigma^{''}\mu^{''}}
=C^{\{\nu\}\beta,\tau\{\nu^{'}\}\beta^{'}\{\nu^{''}\}\beta^{''}}
_{\{\sigma\}\theta\sigma^{'}\sigma^{''},\{\mu\}\phi\mu^{'}\mu^{''}}
\end{equation}
\\
Therefore these CFP are independent of m and n.\\ 
Transforming the Yamanouchi basis of $S(f)$ 
in to the $S(f)\supset S(f_1)\times SU(f_2)$ basis by
\begin{eqnarray}
 \left| \begin{array}{c} \{\nu\} \\ m,\beta\{\sigma\}W_{1}\{\mu\}W_{2} 
\end{array}     \right\rangle 
&=& \sum_{\nu^{''}m^{''}\tau}
 \left( \begin{array}{c} \{\nu\}\\ m \end{array}\right)
 \left|\{\nu\}, \begin{array}{cc}
  \tau\{\nu^{'}\} & \{\nu^{''}\}\\
   m^{'}   & m^{''} \end{array} \right)
\nonumber\\
&&\left|\begin{array}{c}
\{\nu\}\\
\tau\{\nu^{'}\}m^{'}\{\nu^{''}\}m^{''},\beta\{\sigma\}W_{1}\{\mu\}W_{2}
\end{array} \right\rangle 
\end{eqnarray}
and after some algebra one obtains the $SU(mn)\supset SU(m)\times SU(n)$ CFP
 as
\begin{eqnarray}
&C^{\{\nu\}\tau,\beta\{\sigma\}\theta\{\mu\}\phi}
_{\{\nu^\prime\}\beta^\prime\sigma^\prime\mu^\prime,\{\nu^{\prime\prime}\}
\beta^{\prime\prime}
\sigma^{\prime\prime}\mu^{\prime\prime}  }
=\sum_m^{fix. m^\prime} \sum_{m_1m_2m^{\prime\prime}_1m^{\prime\prime}_2} 
C^{\{\nu\}\beta,m}_{\sigma m_1,\mu m_2} 
C^{\{\nu^\prime\}\beta^\prime,m^\prime}_{\sigma^\prime m_1^\prime,
\mu^\prime m_2^\prime}  
C^{\{\nu^{\prime\prime}\}\beta^{\prime\prime},m^{\prime\prime}}_
{\sigma^{\prime\prime} m_1^{\prime\prime},\mu^{\prime\prime}
 m_2^{\prime\prime}}
\nonumber \\
&\left( \begin{array}{c} \{\nu\}\\ m \end{array} 
\left|\{\nu\}, \begin{array}{cc} \tau\{\nu^{\prime}\} & 
\{\nu^{\prime\prime}\}\\
  m^\prime & m^{\prime\prime} \end{array} \right.\right)
\left(\begin{array}{c} \{\sigma\}\\ m_1 \end{array} 
\left|\{\sigma\}, \begin{array}{cc}  \theta\{\sigma^\prime\} 
&\{\sigma^{\prime\prime}\}\\   m_1^\prime  & m_1^{\prime\prime} \end{array} 
\right. \right)
\nonumber \\
&\left(  
\begin{array}{c} \{\mu\}\\ m_2 \end{array} 
\left|\{\mu\}, \begin{array}{cc}
 \phi\{\mu^\prime\} & \{\mu^{\prime\prime}\}\\
  m_2^\prime   & m_2^{\prime\prime} \end{array} \right. 
\right) 
\end{eqnarray} 
where the C on the RHS are the Clebsch-Gordan Coefficients.
\\
This is rather a complex looking expression. It however simplifies
 considerably for the  case of totally antisymmetric irreps 
$\{\nu^{'}\} = \{1^{f_{1}}\} , \{\nu^{''}\} = \{1^{f_{2}}\},
  \{\nu\} = \{1^{f}\}$ \\
whence
\begin{equation}
 C^{\{1^{f}\},\{\sigma\}\theta\{\mu\}\phi}_{\{1^{f_{1}}\}
\{\sigma^{'}\}\{\mu^{'}\},\{1^{f_{2}}\}\{\sigma^{''}\}\{\mu^{''}\}}
=\delta_{\tilde{\sigma}\mu}\delta_{\tilde{\sigma}^{'}\mu^{'}}
\delta_{\tilde{\sigma}^{''}\mu^{''}}\delta_{\theta\sigma}
\left(\frac {h_{\sigma^{'}}h_{\sigma^{''}}}{ h_{\sigma}}\right)^{\frac{1}{2}} 
\end{equation}
where the $h_{\sigma^{\prime}}$, $h_{\sigma^{\prime\prime}}$ and  $h_{\sigma}$, are the 
dimensions of the irreps $\{\sigma^{\prime}\}$,$\{\sigma^{\prime\prime}\}$
 and $\{\sigma\}$ 
of the permutation groups $S(f_{1})$,$S(f_{2})$ and $S(f)$ respectively. 
The last expression is what we will need to calculate the hidden colour 
components of multiquark systems.
\\
The results for the colourless channel are given in the table. For example C
 for the 9-quark state for the state $(56,1)\bigotimes (490,1)
\rightarrow(980,1)$ of $SU(12)\supset SU(6)\bigotimes SU(3)_{c}$ is given
 as
\begin{equation}
 C^{[1^{9}][333]\tilde{[333]}}_{[1^{3}][3][3],[1^{6}][33]\tilde{[33]}}
=\left (\frac {h_{[3]}h_{[33]}}{ h_{[333]}}\right)^{\frac{1}{2}} 
=\sqrt{\frac{5}{42}} 
\end{equation}

Multiplying this with $\sqrt{\frac{1}{5}}$ of $(490,1)$ we get
 $\sqrt{\frac{1}{42}}$
for the colourless channel. Therefore the 9-quark state consists of
 $97.6\%$ hidden colour components ( see Table ).

\vskip 0.5 cm

\begin{table}
\centerline{\bf Table}
\centerline{ Hidden colour components in multiquark systems}
\vskip 0.2 cm

\begin{center}
\begin{tabular}{|c|c|}
\hline
 multiquark configurations & Percentage of hidden colour \\
\hline
$ qq\bar{q}\bar{q}$ (qq in [6])  &  33\\
\hline
$ qq\bar{q}\bar{q}$ (qq in [3])  & 66\\ 
\hline
Pentaquark  $ qqqq\bar{q}$ & 66 \\
\hline
A = 2  $ q^6$ & 80 \\
\hline
A = 3  $ q^9$ & 97.6 \\
\hline
A = 4  $ q^{12} $ & 99.8 \\
\hline

\end{tabular}
\end{center}
\end{table}

\newpage
{\bf HIDDEN COLOUR HYPOTHESIS}
\vskip .5 cm

If colour singlet is so basic to nature, the author feels that the 
allied concept of the hidden colour should not be any less important. 
Hence the author would like to propose here a new hypothesis. 
This actually very simply arises as an extension of the colour singlet 
hypothesis and in a manner with physical manifestations opposite
to those of the colour singlet hypothesis. 
Let us call it the "Hidden Colour Hypothesis". 
As per this hypothesis, in the ground state or at low excitations, 
the creation as an independent entity of a hadron which  
has a large hidden colour part in its wave function would be 
appropriately suppressed. Hence if in a particular configuration in 
space-time a particular hadron has large hidden colour component in 
the state in which it is expected to exist, then
it will be difficult to produce it in the laboratory.
Also if deep inside colour singlet objects, it will manifest as repulsive 
core in the appropriate region.

As an application of this hypothesis let us look at deuteron. 
Most of the time the two nucleons in deuteron are isolated colour 
singlet objects. However
when two such nucleons come together to form a bound system like
deuteron, why do they not have configuration where the two nucleons overlap
strongly in regions of size $ \leq 1 fm $ to form 6-quark bags ? Why is
deuteron such a big and loose system ? The reason has to do with the
structure of the 6-q bags formed had the two nucleons overlapped strongly.
At the centre of mass of the bound system clearly the six quark 
configuration should be important. Hence this $ 80 \% $ hidden
colour part ( see Table ) would prevent the 
two nucleons to come together and overlap
strongly. Therefore the hidden colour would manifest itself as a
short range repulsion in the region $ \leq 1 fm $ in deuteron. So the two
nucleons though bound, stay considerably away from each other.
Hence this new hypothesis is able to explain the basic property of short 
range repulsion in nuclear physics.

The hidden colour concept in spite of being successful has had a rough start.
There have been some claims that hidden colour may not be a 
useful concept as these hidden colour states can be rearranged in terms of
asymptotic colour singlet states. But as discussed in literature \cite{ref2}
the hidden colour concept is not unique
only when the two clusters do not overlap strongly and asymptotically can
be separated out. However when the clusters of 3-q each overlap strongly
so that the relative distance between them goes to zero then the hidden
colour concept becomes relevant and unique \cite{ref2}.
We would like to point out
that indeed, this necessarily is the situation for deuteron discussed
above. Also note that for the ground state the quark configuration is 
$ s^{6} $ given by configuration space representation [6] while $
s^{4}p^{2} $ given by [4] does not come into play as there is not
enough energy to put two quarks into the p-orbital.

Group theoretically the author had earlier obtained the hidden colour
components in 9- and 12-quark systems \cite{ref3, ref4}. 
For the ground state and low energy description of nucleons we
assume that $ SU(2)_{F} $ with u- and d-quarks is required. Hence we
assume that 9- and 12-quarks belong to the totally antisymmetric
representation of the group 
$ SU(12) \supset SU(4)_{SF} \otimes SU(3)_{C} $ where
$ SU(3)_{c} $ is the QCD group and  
$ SU(4)_{SF} \supset SU(2)_{F} \otimes SU(2)_{S} $  where S denotes spin.
Note that up to 12-quarks can sit in the s-state in the group SU(12). The
calculation of the hidden colour components for 9- and 12-quark systems
requires the determination of the coefficients of fractional parentage 
for the group $ SU(12) \supset SU(4) \otimes SU(3) $ \cite{ref3, ref4}
which becomes quite complicated for large number of quarks (see Appendix).
The author found that the hidden colour component \cite{ref3, ref4} 
of the 9-q system is $ 97.6 \% $ while the 12-q
system is $ 99.8 \% $ ie. practically all coloured ( see Table ).

\vskip .6 cm
{\bf A = 3,4 NUCLEI}
\vskip .6 cm

Where would these 9- and 12-quark configurations be relevant in nuclear
physics ? The A=3,4 nuclei 
$ ^{3}H $, $ ^{3}He $ and $ ^{4}He $ have sizes of 
1.7 fm, 1.88 fm and 1.674 fm respectively. Given the fact that each
nucleon is itself a rather diffuse object, quite clearly in a size
$ \leq 1 fm $ at the centre of these nuclei the 3 or 4 nucleons would
overlap strongly. As the corresponding 9- and 12-q are predominantly
hidden colour, there would be an effective repulsion at the centre keeping
the 3 or 4 nucleons away from the centre as per the hidden colour 
hypothesis above. Hence it was predicted by the
author \cite{ref3,ref4} that there should be a hole at the centre of 
$ ^{3}H $, $ ^{3}He $ and $ ^{4}He $.
And indeed, this is what is found through electron scattering. 
Hence the hole, ie. significant depression in the central density of 
$ ^{3}H $, $ ^{3}He $ and $ ^{4}He $ is a signature of 
the hidden colour hypothesis in this ground state property.

\newpage
{\bf NUSOSPIN}
\vskip .5 cm

The author has used this concept of the role of hidden colour in
A=3,4 nuclei to predict triton clustering in nuclei \cite{ref5}.
Equally significant is the suggestion of a new symmetry in the same 
neutron rich nuclei \cite{ref6}. Herein 
triton ("t")  ${^{3}_{1} H_{2}}$
helion ("h")  ${^{3}_{2} He_{1}}$ are treated as
fundamental representations of a new symmetry group  $SU(2)_{\cal A}$
\cite{ref6} which has been christened as "nusospin symmetry". 
Note that basically this nusospin symmetry arises due to the application 
of the hidden colour hypothesis.
Clearly the fact that  ${^{3Z}_Z} A _{2Z}$ nuclei 
are made up of Z number of tritons leads to new stability for them.
The author has found extra-ordinary stability manifested 
by the separation energy data for the proton and neutron pairs 
(Z,N): (6,12), (8,16), (10,20), (11,22) and (12,24) \cite{ref7}. 
This also hints at the possibility of neutron stars.

\vskip .5 cm
{\bf EXOTIC HADRONS}
\vskip .5 cm

Now let us look at the significance of hidden colour for the exotic hadrons 
like the diquonia, pentaquarks, dibaryons etc. The recent claims of the 
discovery of the pentaquark has been much in the news. However even 
the discoverers of this exotic states ask for caution in acceptance of 
the claim. This should be seen in the light of the lack of exotic states 
in spite of intense searches for these in laboratories all over the 
world these last few decades. 
Once in a while a claim is made. But mostly it does not withstand 
closer scrutiny as none of these are starred objects in the Particle Data 
Tables! This is by and large the story of the exotic baryons. 
So one may ask as why, if QCD demands their existence, is it that still 
we do not have a definite candidate for the exotic hadron as of now.

The answer is provided by the new hidden colour hypothesis proposed here.
We have already seen that inside nuclei these hidden colour components 
can manifest themselves over a short range near the centre of mass of the 
nucleus and there they provide for short range repulsion in two, three 
nucleon and four nucleon systems. From the table  we notice that these 
exotic hadrons also have large hidden colour parts and so if we try to to 
create these in these configurations, it will be resisted. 
Also in the ground state we feel that these configuration should be 
significant.  Hence in short, the hidden colour hypothesis says that it 
will be difficult to create these exotic hadrons and thereby explaining 
their reluctance to show up as of now. 

It should be pointed out that major support for the colour singlet 
hypothesis comes from the fact that as of now we have not found in free 
state any coloured object. So now the fact that as of now we have not 
found a definite candidate for exotic hadrons, should be taken as a 
proof or strong motivation for the hidden colour hypothesis.
One exception may be the pentaquark. But in that case as per the table
in one particular case the hidden colour part is only 33 per cent.
The small resistance may be bypassed by some ingenious experimentation.
However one possible prediction of this hypothesis is that as the 
individual multiquarks have large hidden colour parts within them 
it may be that two or more multiquarks of the same kind may like 
to form a composite as then they may have many more ways of forming a 
colour singlet hadron. So perhaps the experimantalists should look for 
these objects as clusters of two or more of the same kind of multiquarks.


\vskip .7 cm

\begin{thebibliography}{99}

\bibitem{ref1}
V. A. Matveev and P. Sorba, 
{\it Lett. al Nuovo Cim.} {\bf 20} (1977) 435

\bibitem{ref8}
J. Q. Chen, J. Math. Phys. {\bf 22} (1981) 1

\bibitem{ref2}
P. Gonzalez and V. Vento, 
{\it Il Nuovo Cimento} {\bf 106 A} (1992) 795

\bibitem{ref3}
A. Abbas, {\it Phys.Letts.} {\bf 167 B} (1986) 150

\bibitem{ref4}
A. Abbas, {\it Prog. Part. Nucl. Phys.} {\bf 20} (1988) 181 

\bibitem{ref5}
A. Abbas, Mod. Phys. Lett. {\bf A 16} (2001) 755

\bibitem{ref6}
A. Abbas, Mod. Phys. Lett. {\bf A 19} (2004) 2365

\bibitem{ref7}
A. Abbas, Mod. Phys. Lett. {\bf A 20} (2005) 2553

\end{thebibliography}
\end{document}